\documentclass[12pt]{iopart}

\usepackage{amssymb}
\usepackage{graphicx}
\usepackage{color}
\usepackage{hyperref}
\usepackage{cite}

\begin{document}

\title{Vlasov simulation of laser-driven shock acceleration and ion turbulence}
\author{A. Grassi$^{1,2,3}$, L. Fedeli$^{1,3}$, A.Sgattoni$^{3}$, A. Macchi$^{1,3}$}

\address{$^1$Dipartimento di Fisica Enrico Fermi, Universit\`a di Pisa, Largo Bruno Pontecorvo 3, I-56127 Pisa, Italy}
\address{$^2$LULI, Universit\'{e} Pierre et Marie Curie -\'{E}cole Polytechnique, Paris, France}
\address{$^3$Istituto Nazionale di Ottica, Consiglio Nazionale delle Ricerche (CNR/INO), u.o.s. Adriano Gozzini, Pisa, Italy}


\begin{abstract}
We present a Vlasov, i.e. a kinetic Eulerian simulation study of nonlinear collisionless ion-acoustic shocks and solitons excited by an intense laser interacting with an overdense plasma. The use of the Vlasov code avoids problems with low particle statistics and allows a validation of particle-in-cell results. A simple, original correction to the splitting method for the numerical integration of the Vlasov equation has been implemented in order to ensure the charge conservation in the relativistic regime. We show that the ion distribution is affected by the development of a turbulence driven by the relativistic ``fast'' electron bunches generated at the laser-plasma interaction surface. This leads to the onset of ion reflection at the shock front in an initially cold plasma where only soliton solutions without ion reflection are expected to propagate. We give a simple analytical model to describe the onset of the turbulence as a nonlinear coupling of the ion density with the fast electron currents, taking the pulsed nature of the relativistic electron bunches into account. 
\end{abstract}


\submitto{\PPCF}

\section{Introduction}

The production of high energy protons by laser-driven acceleration is currently a subject of active research. Several experiments have demonstrated the generation of multi-MeV proton beams in a wide range of laser and target parameters \cite{Review,Review_daido,MacchiPPCF,Fernandez}. Different mechanisms have been studied in order to optimize and control the main characteristics of the proton beam for the foreseen applications. 

Among the various ion acceleration mechanisms, here we focus on collisionless shock acceleration (CSA). The simplest CSA scenario may be briefly described as follows. Collisionless shock waves can be excited at the laser-plasma interaction surface due to the combination of radiation pressure-driven pushing and steepening of density profile and of electron heating up to high temperatures. In the electrostatic regime (to which we restrict ourselves in the present paper) the shock waves are characterized by an electrostatic potential jump at the shock front, moving with velocity $V_s$.
Some ions can be accelerated from rest by reflection from the shock front as a moving wall, with a velocity gain of $2V_s$, provided that the ions have a sufficient velocity along the direction of the shock propagation so that they cannot overcome the potential barrier. 
For a plasma with electron temperature $T_e$, the shock velocity is $V_s=Mc_s$ where $M>1$ is the Mach number and $c_s=(ZT_e/m_i)^{1/2}$ is the ion acoustic velocity with $m_i=Am_p$ is the ion mass, $m_p$ the proton mass, $Z$ and $A$ the ion charge and mass numbers. 
At laser irradiances $I\lambda^2 >10^{18}~\mbox{W cm}^{-2}\mu\mbox{m}$ (with $I$ and $\lambda$ the laser intensity and wavelength, respectively) ``fast'' high--energy electrons are generated with effective values of $T_e$ of several MeVs. Thus, reflection from the shock front may generate multi--MeV ions with typical energy ${\cal E}_i=m_i(2V_s)^2/2=2M^2ZT_e$, and a monoenergetic spectrum as far as the shock velocity remains constant.
 
The CSA mechanism has been explored previously via particle-in-cell (PIC) simulations \cite{DenavitPRL,SilvaPRL,HePRE07,macchiPRE12,FiuzaPOP,FiuzaPRL_es,liseykinaJPP15} and in recent laboratory experiments using linearly polarized, $\mathrm{CO_2}$ laser ($\lambda=10~\mu\mbox{m}$) pulses \cite{haberbergerNP12,chenSPIE15,trescaPRL15,ZhangPOP15} and gas jet targets which were slightly overdense for the laser pulse, i.e. having an electron density $n_e \gtrsim n_c=4\pi m_ec^2/(e^2\lambda^2)$, the cut-off (or ``critical'') density.
In the experiment by Haberberger et al.\cite{haberbergerNP12} a narrow monoenergetic peak corresponding to $\sim 20$~MeV protons was observed using a linearly polarized train of pulses with duration of a few picoseconds.
While the possibility to accelerate monoenergetic protons is appealing, the proton flux observed in Ref.\cite{haberbergerNP12} is relatively low ($\simeq 10^7~\mbox{MeV}^{-1}\mbox{sr}^{-2}$) so that the population of accelerated protons would be hardly resolved in a PIC simulation, unless the number of particles per cell is drastically increased up to values that would be computationally challenging for multi-dimensional simulations. 
Moreover, PIC simulations of CSA have shown the crucial role of the background ion temperature on shock formation and particle reflection \cite{macchiPRE12}, so that the high resolution of the low-density tail and of the non-thermal component in the ion distribution function is important. This issue suggests that PIC simulations should be tested when possible against Eulerian ``Vlasov'' simulations which are free from effects of low particle statistics and fluctuations, at the cost of higher computational requirements. 

In this paper, we studied CSA using a Vlasov code and comparing the results to those from PIC simulations. The numerical approach is described in \sref{Codice}. The Vlasov code includes a simple method to adapt the splitting algorithm to the relativistic case without violating mass conservation, as described in the Appendix. 
Our results include the observation of acceleration at the shock front also in the case with no thermal spread for the initial ion distribution function ($T_{i}\simeq0$) where the theory (resumed in \sref{Theory}) predicts only the generation of a solitary wave, i.e. an ion-acoustic soliton, without reflection from the shock front. This observation is related to the development of a sort of ion density turbulence in the upstream region ahead of the shock front, locally broadening the ion velocity space and allowing for ion reflection. In \sref{Modello} we give a simple analytical model in which, on the basis of the observations, we relate the ion turbulence to the pulsed nature of the fast electron bunches generated by the laser-plasma interaction. 

\section{Background theory}
\label{Theory}

The formation of a collisionless electrostatic shock solution is not possible in principle within the fluid description, where only soliton solutions appear as the result of a balance between nonlinearity and dispersion. A shock wave solution may exist in presence of kinetic effects acting as an effective dissipation, breaking the symmetry of the soliton \cite{Sagdeev}. In this framework, ion acceleration due to reflection from the shock front may be a mechanism inherent to the collisionless shock formation rather than a consequence of it.

A theory of collisionless nonlinear ion-acoustic shocks and solitons is resumed in Refs.\cite{Sagdeev,Tidman} where the electrons are assumed to be non-relativistic and in Boltzmann equilibrium. In the shock front rest frame, moving at velocity $V_s$ with respect to the laboratory frame, the reflection occurs for all the ions whose kinetic energy does not exceed the height of electrostatic potential barrier $\phi_{m}$. All the ions in the high energy tail of the distribution function with a velocity component $v_{i}$ along the shock propagation direction in the laboratory frame such that
\begin{equation}
v_{i}>V_{s}-\sqrt{\frac{2e\phi_{m}}{m_{i}}}
\label{eq:minimun_vi}
\end{equation}
will get reflected from the shock. Thus, a spread in the initial velocity distribution function is required, otherwise no reflection occurs and only soliton solutions are possible. The number of reflected ions should be however small enough to avoid loading of the shock, e.g. excessive energy transfer to the accelerated particles \cite{macchiPRE12}. This makes clear the importance of the initial ion distribution in order to predict the development and lifetime of a soliton or a shock wave and the fraction of accelerated particles.

Heuristically, the formation of high-speed shocks or solitons in intense laser interaction with overdense plasmas involves two requirements: the electrons must be heated to high temperatures to allow shock/soliton propagation, and a strong initial density perturbation must be driven. For what concerns the first requirement, a linearly polarized pulse generates a fast electron population with a typical kinetic energy 
\begin{equation}
{\cal E}_p=m_ec^2\left(\left(1+a_0^2/2)^{1/2}-1\right)\right) \; ,
\label{eq:ponderomotive}
\end{equation}
where $a_0=(2I/m_ec^3n_c)^{1/2}=0.85(I\lambda^2/10^{18}~\mbox{W cm}^{-2}\mu\mbox{m})^{1/2}$ is the peak dimensionless amplitude of the laser field.  If we roughly assume the fast electrons to have a Boltzmann distribution with temperature $T_f \simeq {\cal E}_p$, then the fast electrons can sustain the propagation of nonlinear waves characterised by the ion-acoustic velocity $c_s=(ZT_f/m_i)^{1/2}$. Recirculation of the fast electrons through the target is also necessary to produce a uniform temperature, which may introduce a dependence upon the pulse duration and target thickness.

The driving perturbation can be provided by the radiation pressure of the laser pulse $P_L=(1+R)I/c$ (with $R\leq 1$ the reflection coefficient) pushing the plasma surface as a piston with an average velocity $v_{\mbox{\tiny HB}}$, commonly known as the hole boring (HB) velocity. The expression of $v_{\mbox{\tiny HB}}$  can be obtained either via a momentum balance equation \cite{robinsonPPCF09} or via a dynamic model \cite{macchiPRL05} as (for $v_{\mbox{\tiny HB}}\ll c$)
\begin{equation}
v_{\mbox{\tiny HB}} = \left(\frac{P_L}{2m_ic^2}\right)^{1/2}=\left(\frac{1+R}{2}\frac{Z}{A}\frac{m_e}{m_p}\frac{n_c}{n_e}\right)^{1/2}\frac{a_0}{\sqrt{2}} \; .
\label{eq:Vp}
\end{equation} 
The dynamic model shows that a fraction of ions is accelerated up to a velocity $2v_{\mbox{\tiny HB}}$ (following wave breaking of the density profile \cite{macchiPRL05}). The generation of such ions  is also needed for momentum conservation \cite{robinsonPPCF09,schlegelPoP09}.  Density perturbations may detach from the surface and propagate with supersonic speed ($M>1$) into the plasma when $v_{\mbox{\tiny HB}} \gtrsim c_s$. 

A collisionless shock may also be generated due to the nonlinear evolution of instabilities, which in the multi-dimensional case include electromagnetic counterstreaming (or ``Weibel'') instabilities driven by fast electrons \cite{FiuzaPRL-Weibel}. In Ref.\cite{haberbergerNP12} the authors advocate some driving mechanism different by radiation pressure in order to account for a shock velocity $V_s\gg v_{\mbox{\tiny HB}}$. In the present paper we restrict to one-dimensional (1D) geometry and focus on radiation pressure-driven shocks.

Notice that in a cold plasma, i.e. in the absence of fast electrons, a ``true'' shock or soliton wave may not form and detach from the surface. In this case, only the above mentioned ion population with velocity $\simeq 2v_{\mbox{\tiny HB}}$ is observed; such ions move into the plasma ballistically, as charge-neutralized bunches \cite{macchiPRL05}. This is the so-called HB acceleration regime. Although one may consider such ions as ``reflected'' from the laser-driven piston, so that the process sounds quite similar to CSA, there are basic differences between CSA and HB acceleration, which is favored if circular polarization (and normal incidence) of the laser pulse is used instead of linear polarization since in the former case the fast electron generation is strongly reduced \cite{macchiPRL05}. While the shocks driven in CSA accelerate ions along the propagation in the plasma, HB occurs only at the plasma surface and during the laser pulse action. In addition, the number of ions accelerated by HB is larger (the density of ion bunches may be close to the background value). This might explain why the flux of detected protons reported in Ref.\cite{haberbergerNP12} is some five orders of magnitude lower than that observed with a similar experimental set-up, but using circular polarization \cite{palmerPRL11}. In this case the proton spectrum shows a broader peak at the lower energy $\simeq 1~\mbox{MeV}$ which is fairly consistent with HB theory. In the case of linear polarization, HB and CSA may both occur.

\section{Numerical method and simulation set-up }
\label{Codice}

We developed a Vlasov code that provides a solution of the 1D Vlasov-Maxwell system of equations within a completely Eulerian approach. Once the distribution function of each species is discretized on a fixed grid in the real space and in the momentum space, the code performs a direct integration of the Vlasov equation. The Vlasov equation is coupled with Maxwell\textquoteright{}s equations for self-consistent electromagnetic fields, generated by the charge densities and currents of the particle species. These latter quantities are defined as the momenta of the distribution function and are calculated over all the Eulerian grid cells. In the absence of a magnetic field component along the propagation direction ($B_{x}=0$), the Vlasov equation can be reduced to a 1D1P geometry exploiting the conservation of transverse canonical momentum $\vec{\Pi}_{\perp}=\vec{p}_{\perp}+q\vec{A}_{\perp}$, where $\vec{A}_{\perp}$ denotes the transverse component of the vector potential. Thus for a particle species with mass $m$ and charge $q$, the evolution of the reduced one-dimensional distribution function $f=f(x,p_x,t)$ is described by 
\begin{equation}
\frac{\partial f}{\partial t}+\frac{p_{x}}{m\gamma}\frac{\partial f}{\partial x}+q\left(E_{x}-\frac{q}{2m\gamma}\frac{\partial|\vec{A}_{\perp}(x,t)|^{2}}{\partial x}\right)\frac{\partial f}{\partial p_{x}}=0\label{eq:Vlasov}
\end{equation}
For the numerical integration of eq.(\ref{eq:Vlasov}), the Time Splitting Scheme \cite{Cheng1976330} and the Positive and Flux Conservative Method \cite{Filbet2001166} have been employed along with an original method providing the exact mass conservation in the relativistic regime (see \ref{sec:Appendix}). Violation of mass conservation was a known drawback of the use of splitting schemes in the relativistic regime \cite{Huot2003512}. In order to ensure mass conservation
quite complex methods have been previously proposed \cite{Huot2003512,Elkina2006862}. Our method is much simpler than others, but apparently effective. 

In the next section, we report the results of a simulation with an electron-proton plasma of length $L=10\:\lambda$ having a steep density profile with
a linear rising and falling ramp of length $0.2\:\lambda$, where $\lambda$
is the laser wavelength. The plasma plateau density is set to $n_{i}=n_{e}=2.0\:n_{c}$. The laser pulse has linear polarization, peak amplitude $a_{0}=2.0$ and duration $\tau=60\: T$, with $T$ the laser period. The temporal profile has a $\mathrm{{sin}}^{2}$-like rising and falling ramp of one period length, and $58\: T$ of constant plateau. The temporal and spatial resolution is set to $\Delta x=c\Delta t={\lambda}/{10^3}$. 
{
The size of the simulation box is $L_x=16\:\lambda$. 
In the momentum grid the resolution is $\Delta p_{x,e}=0.05\: m_{e}c$ and $\Delta p_{x,i}=0.045\: m_{e}c$, respectively for electrons and ions. The distribution function, for each species, has a Gaussian shape in momentum space with initial temperatures of $\: T_{e}=5\:\:\mathrm{{KeV}\:}$ and $T{}_{i}=1\:\ \mathrm{{eV}}$. A lower electron temperature would require a much greater computational effort, and as soon as the electrons are heated up by the laser-plasma interaction, the spread of the distribution function becomes much higher than the initial distribution. 
}
A slightly higher resolution is chosen for the ion momentum with respect to the electrons, in order to resolve the initial spread of the ion distribution function that is initialized with a temperature much lower than the electron distribution. Open boundaries are used for the fields and reflecting boundaries for the distribution function. We verified that the density of particle reflected at the boundaries is small enough to have a negligible role on the physical results.

\section{Simulation results}
\label{Simulazioni}

\subsection{Vlasov simulation of shock generation}
\begin{figure}
\centering{}
\includegraphics[width=1.0\textwidth]{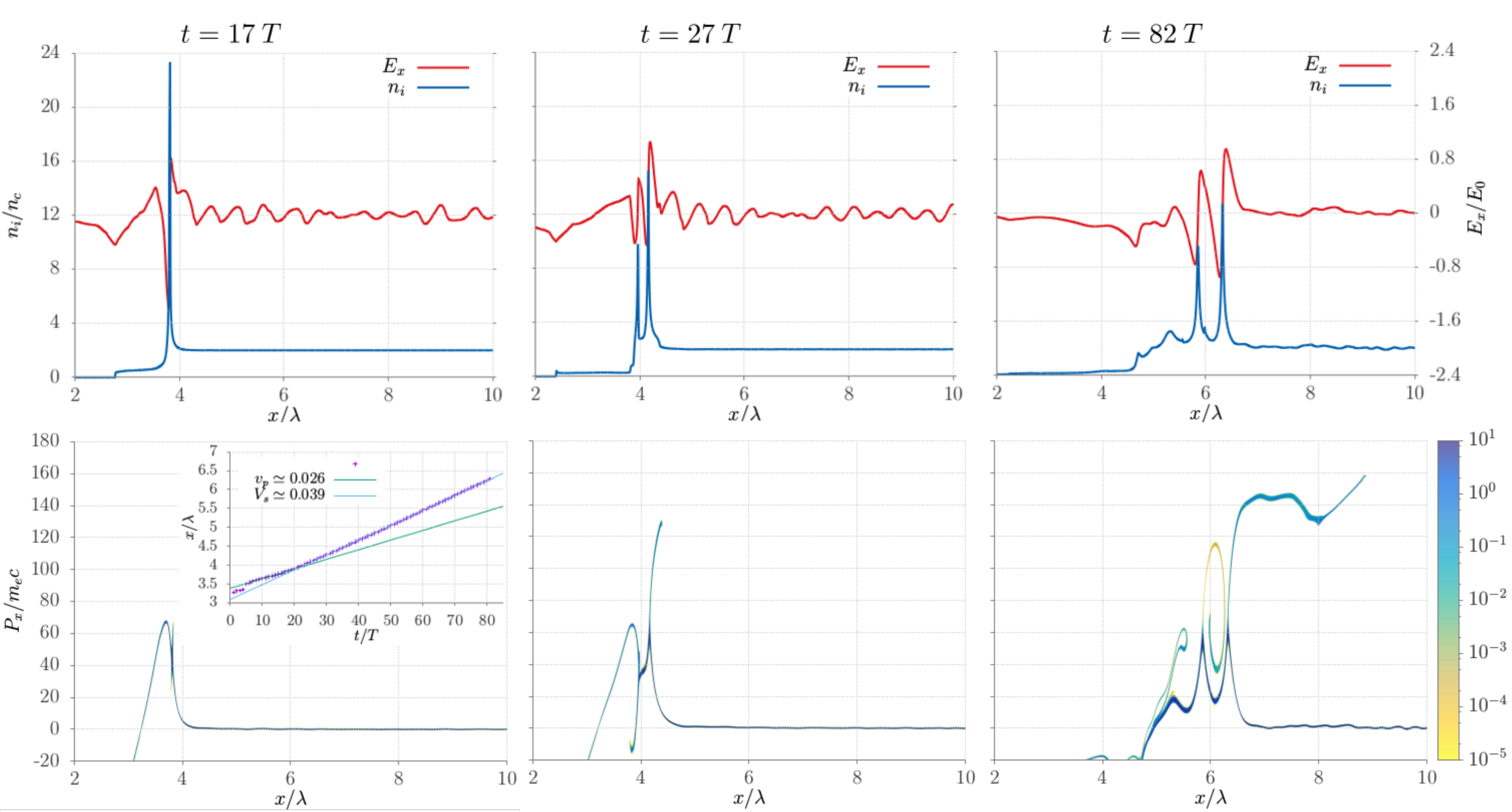}
\caption{\label{PSE_PSI_NIEx}\label{PeakVelocity}
Ion density and longitudinal electric field (in units of $E_{0}={m_{e}c\omega}/{e}$) (top row) and the ion distribution function in the phase space $x-P_x$ (bottom row) at time $t=17,\:27,\:82\:$ in units of the laser pulse period $T=\lambda/c=2\pi/\omega$. 
The inset shows the position vs. time of the rightmost density peak, which is fitted by two linear functions.
}
\end{figure}

\begin{figure}
\centering{}
\includegraphics[width=1.0\textwidth]{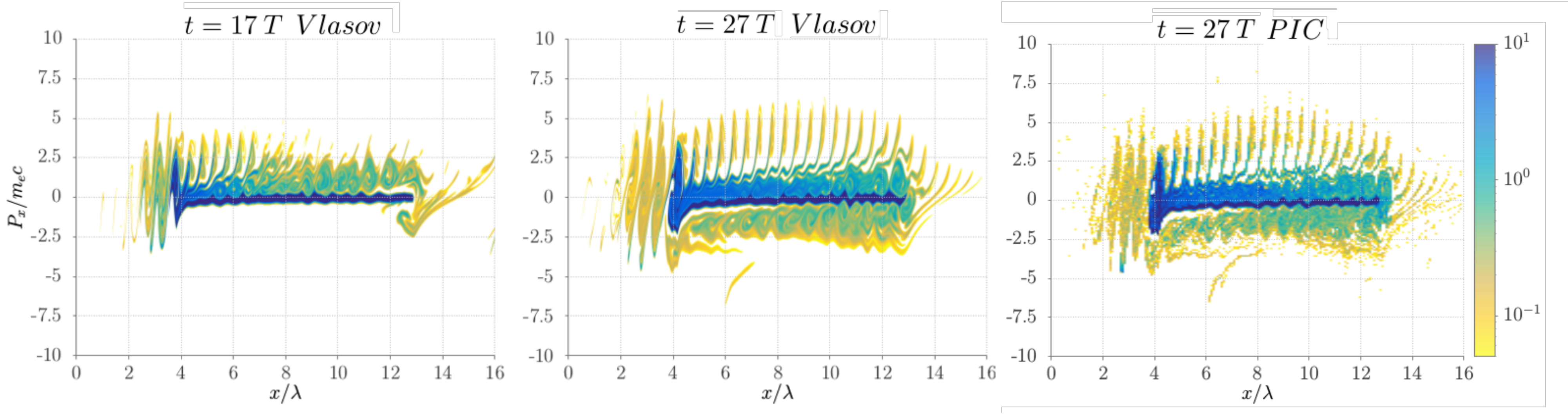}
\caption{\label{PSE}
Electron distribution function in the phase space $x-P_x$ at times $t=17T$ and $t=27T$ for the Vlasov simulation, and at times $t=27T$ for the PIC simulation.
}
\end{figure}

\Fref{PSE_PSI_NIEx} shows snapshots of the ion density $n_i$, the electric field $E_x$, and the ion phase space at different times. The incoming laser pulse propagates from left to right.
At the first time shown in \fref{PSE_PSI_NIEx} ($t=17T$), the sharp density peak generated by the radiation pressure push is apparent. The density peak moves with a velocity of $v_p\simeq 0.026c$, as given by the fit in Fig.\ref{PeakVelocity}, which is slightly higher than the value of $v_{\mbox{\tiny HB}} \simeq 0.023$ (since $R \simeq 0.9$ in the simulation) predicted by Eq.(\ref{eq:Vp}). 
Our explanation is that, in this regime (low density and linear polarization) a significant number of electrons are dragged on the vacuum side, as it is apparent from the phase space plot of the electrons in \fref{PSE}. This causes some ions to be accelerated in the backward direction, i.e. towards the incoming laser pulse, as observed in the ion phase space plot (\fref{PSE_PSI_NIEx}). Since the total momentum flux from the laser to the plasma is always given by $P_L$ and it is in the forward direction, the ions accelerated forward must get an extra boost in addition to that leading to (\ref{eq:Vp}). In other words, fast electrons produce a sort of ``ablation pressure'' which adds up to the laser radiation pressure.

The phase space plot at $t=17T$ in \fref{PSE_PSI_NIEx} shows that the ion density profile is undergoing wavebreaking, which at $t=27T$ has produced to the typical ``{\sf X}''-type phase space structure \cite{forslundPF75,macchiPPCF09,macchiCRP09} with further acceleration of some ions which cross the breaking point \cite{macchiPRL05}.
Correspondingly, the density peak appears to split up in two peaks, the second moving at constant velocity of $V_s=0.039c$ as also shown by the fit in \fref{PeakVelocity}. At later times, an ambipolar electric field structure is evident around the density peak, while a ``pencil'' of ions reflected at velocity $\simeq 2V_s$ appears in the phase space. We thus identify the fast density peak as the front of a radiation pressure-driven shock.    
A fit on the electron spectrum gives a fast electron temperature of $\sim 0.84m_ec^2$ (to be compared with ${\cal E}_p=0.73m_ec^2$ from Eq.(\ref{eq:ponderomotive}), which yields $c_s \simeq 0.021$ and a Mach number $M=V_s/c_s \simeq 1.9$.

The high energy or ``fast'' electrons produced by the interaction of the high-intensity laser with the front surface of the target propagate through the bulk as small-duration bunches (see the electron phase space plot in \fref{PSE}).
When the electrons reach the rear side of the plasma they produce a negative charged sheath. Because of the electric field generated by the expanding sheath at rear side, most of the electrons cannot escape and carry on recirculating across the target. The recirculation leads to a spread of fast electrons all across the target. As shown in \fref{PSE} the first recirculating electrons reach back the front side around $t=27T$, later than the formation of the shock front. The latter keeps a constant velocity also at later times, as shown in \fref{PeakVelocity}. Thus, we conclude that the shock is driven by the radiation pressure action and the related wavebreaking, as the onset of fast electron recirculation does not affect the velocity of the shock. 

\subsection{Ion acceleration and turbulence}
The density and field profile at $t=82T$ in Fig.\ref{PSE_PSI_NIEx} show two prominent density peaks associated to ambipolar field structures. A third, much weaker structure is also visible. The propagation velocity of the second density peak is slower than the first one. This latter observation may suggest that the structures may be better described as two solitary waves, i.e. ion-acoustic solitons, rather than a coherent shock wave. Indeed, as predicted by theory \cite{Sagdeev,Tidman}, in the case of cold ion population we actually do not expect a shock wave but a soliton solution to propagate. For this solution the propagation velocity increases with the amplitude of the electrostatic potential associated with the density peak. However, the ion phase space plot for $t=82T$ in \fref{PSE_PSI_NIEx} clearly shows the presence of ions reflected from the first propagating structure, which is a characteristic feature of shock solutions \cite{Sagdeev,Tidman} (while the ``curly'' feature behind the front corresponds to ions trapped between the two structures). 

\begin{figure}
\centering{}
\includegraphics[width=0.4\textwidth]{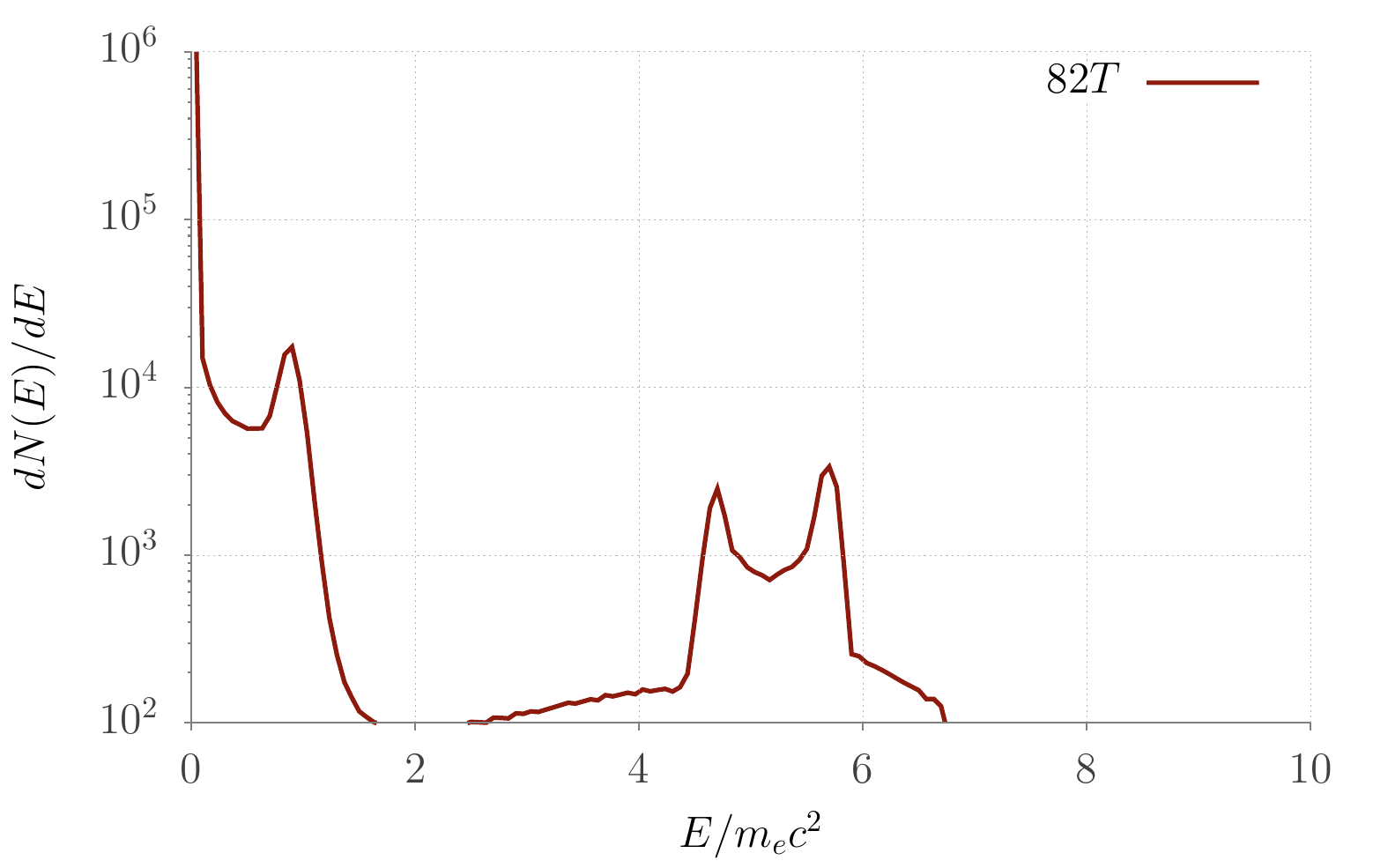}
\includegraphics[width=0.4\textwidth]{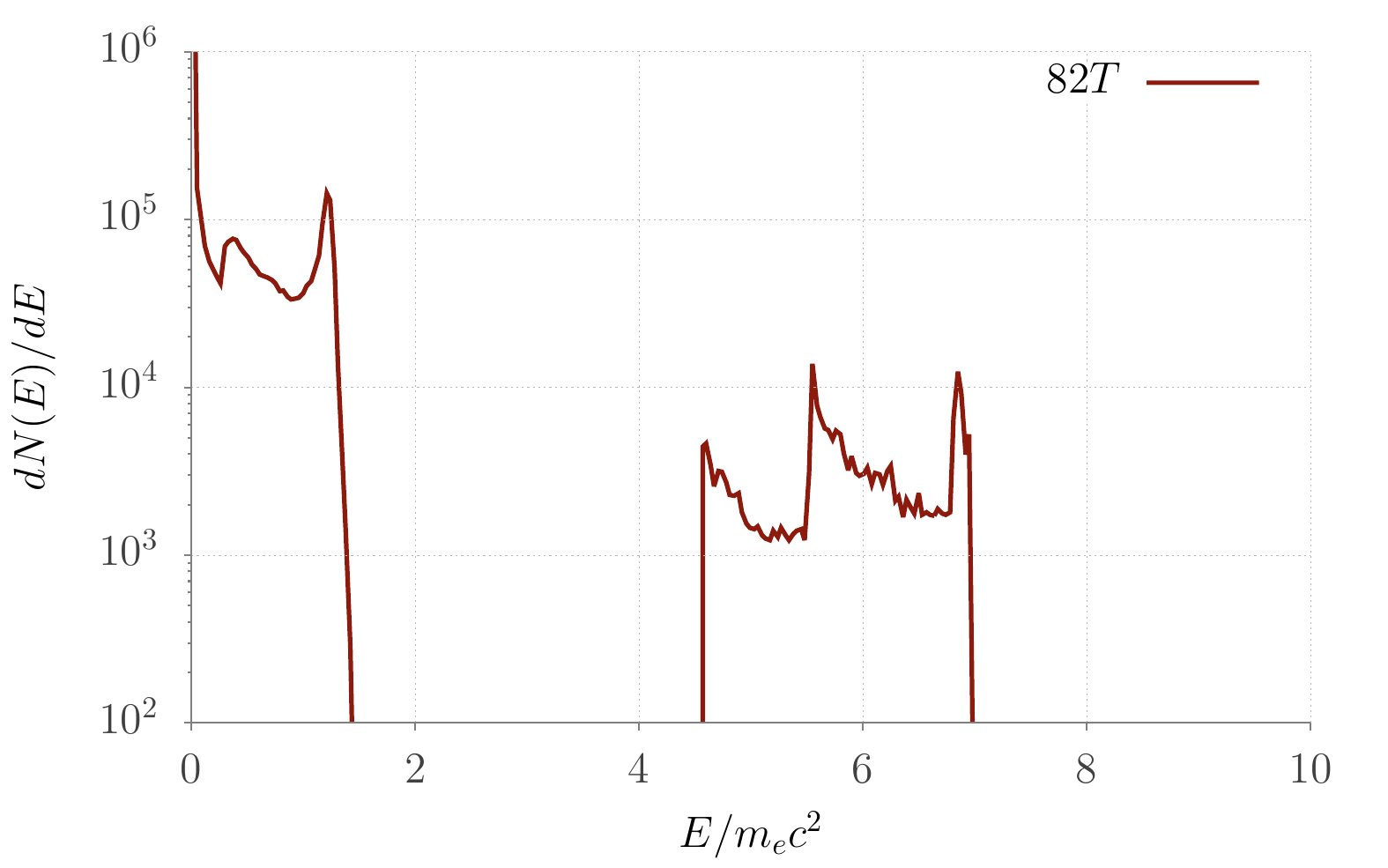}
\caption{\label{fig:EnergySpectrum}Ion energy spectrum at $t=82\: T$ in the $x$-range $6.3\lambda\sim 10 \lambda$ for the Vlasov (left) and the PIC (right) simulations, respectively.}
\end{figure}

\Fref{fig:EnergySpectrum} shows the ion energy spectrum at time $t=82\: T$, which has peaks at energy ${\cal E}\simeq4.7\: m_{e}c^{2}$ and $E\simeq5.7\: m_{e}c^{2}\simeq2.9\:\mathrm{MeV}$ with energy spread $\sim 2\%$. 
The peak at higher energy corresponds to the plateau with momentum $P_{x}\simeq145\: m_{e}c$ in the phase space (\fref{PSE_PSI_NIEx}). This is very close to twice the propagation momentum of the first density peak $P\simeq72.1\: m_{e}c$, as expected for the reflection from a shock front.

The second energy peak around ${\cal E}\simeq4.7\: m_{e}c^{2}$ corresponds to a group of ions which are ahead of the first density peak at $x\simeq8\:\lambda$ and have thus been generated previously, in an early stage of radiation pressure pushing and shock formation and before the stable shock propagation regime. Overall, the ions contained in the two spectral peaks contain a fraction $\lesssim 10^{-3}$ of the laser pulse energy.

In order to estimate the number of reflected particles according to eq.(\ref{eq:minimun_vi}), we measure the value $\phi_m$ of the electrostatic potential jump around the shock front, i.e. the fastest density peak. For the observed value $e\phi_m \simeq 1.3 \:m_ec^2$ the reflection condition (\ref{eq:minimun_vi}) is fulfilled by the ions with velocity $v_i\simeq 70\:v_{th}$, where $v_{th}$ is the thermal velocity corresponding to the initial ion temperature $T_{i}=1\:\mathrm{eV}$. In such region of the phase space the tails of the distribution function are completely negligible ($\sim \mbox{e}^{-70^2} $), thus we should not expect to observe reflected particles, in contrast to what is obtained from the simulation.

\begin{figure}
\centering{}
\includegraphics[width=0.4\columnwidth]{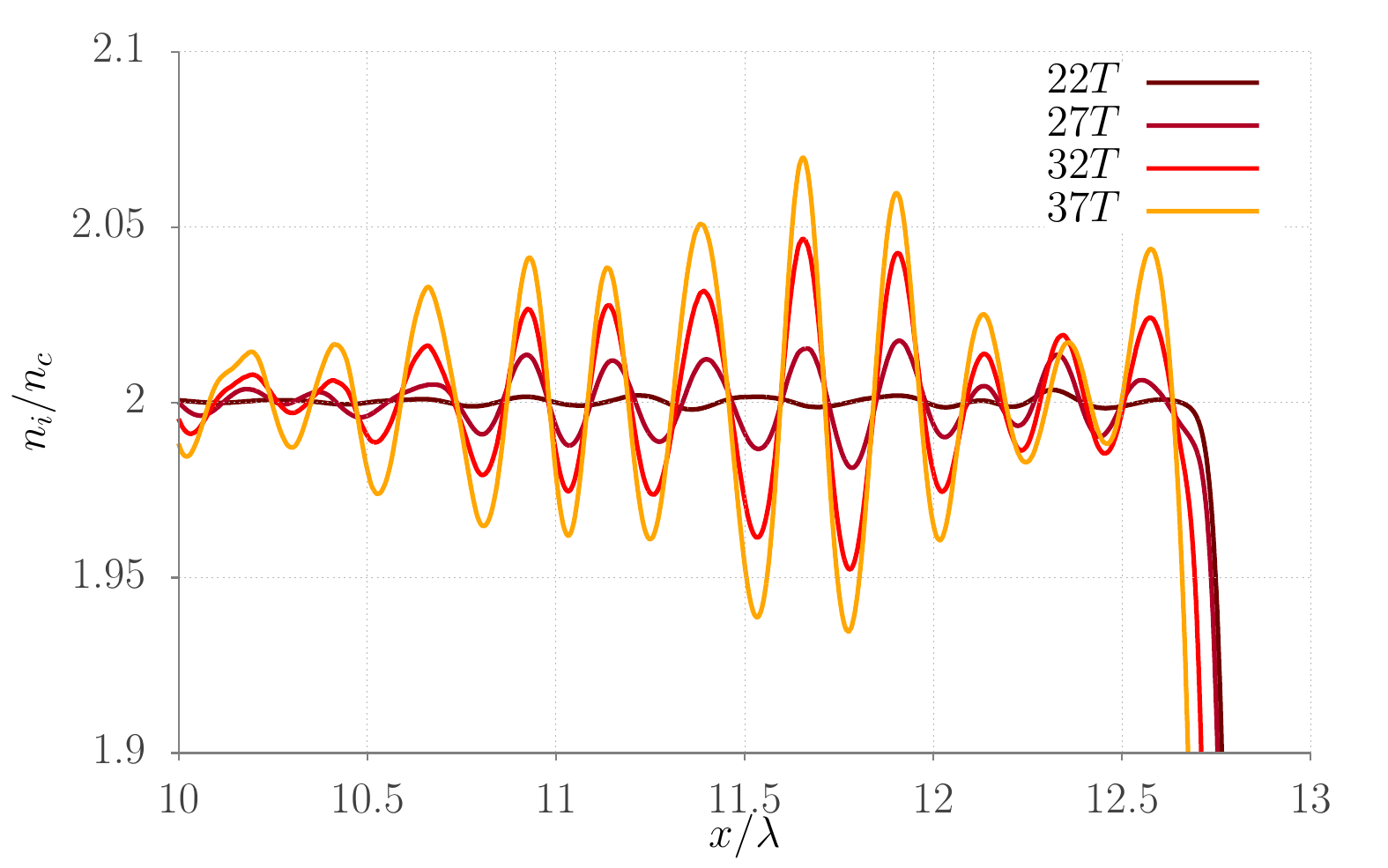}
\includegraphics[width=0.4\columnwidth]{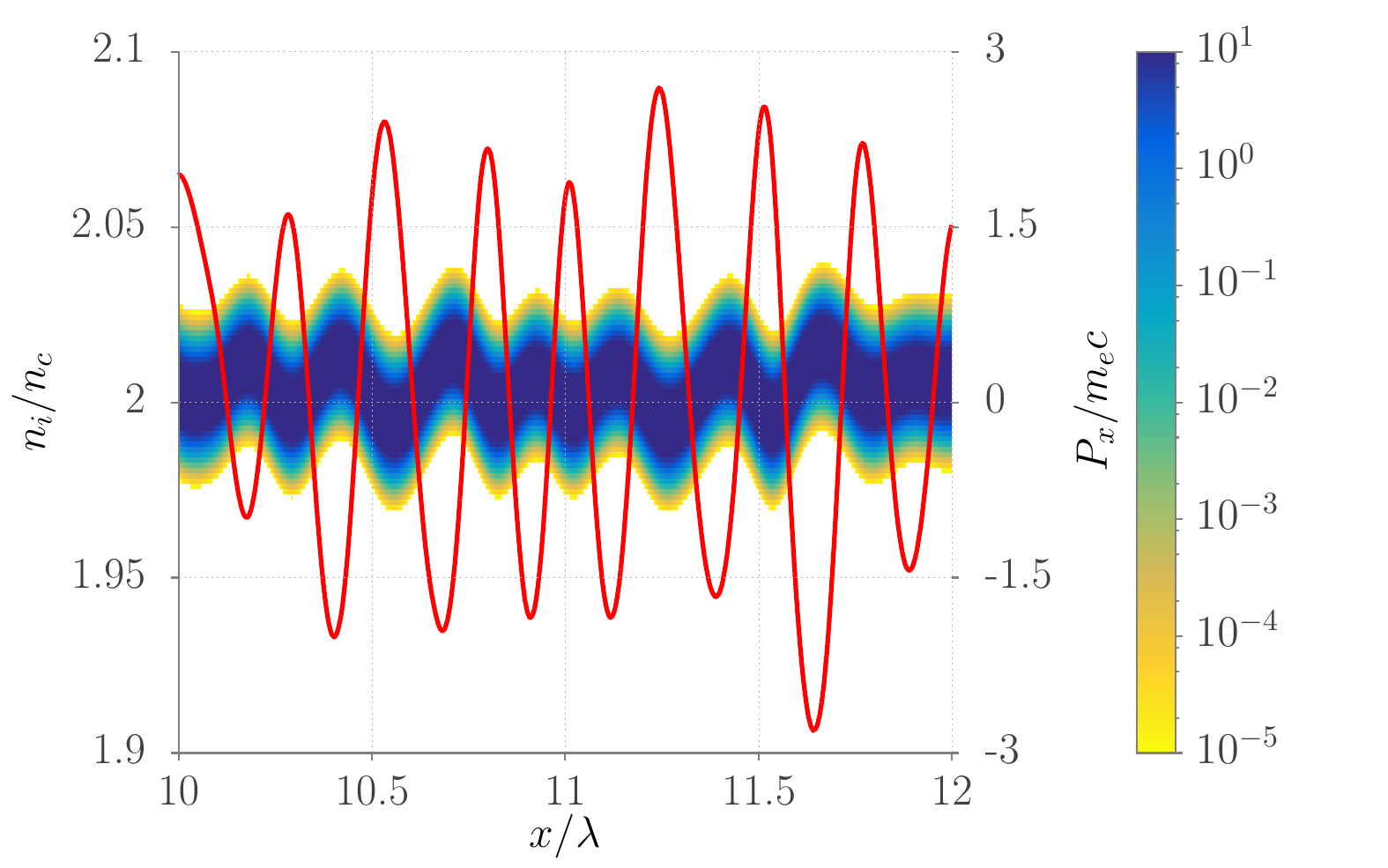}
\caption{\label{fig:Growing-oscillations-of}Left frame: growing oscillations of the ion density showing a spatial period equal to $\lambda/4$, with $\lambda$ the laser wavelength. Right frame: ion distribution function in the phase space $x-P_x$ and ion density (red line) at $t=67\: T$}
\end{figure}

The presence of accelerated particles is justified by the growth of a perturbation in the ion density in the upstream region, where the shock has not yet propagated, as shown in \fref{fig:Growing-oscillations-of}~a). In correspondence with the fluctuations of the ion density, an oscillation of the ion distribution  function is observed in the phase space, see \fref{fig:Growing-oscillations-of}~b).
The oscillations of the ion velocity produce a spread of the distribution function and allow a fraction of ions to exceed the threshold value of Eq.(\ref{eq:minimun_vi}), so that they get reflected by the shock front.  The turbulence in front of the first ion density peak leads to a variable quantity of reflected ions with time.

In order to characterize the ion density perturbation we notice that it starts growing at the rear side of the plasma when the fast electrons have passed through this region, after being pushed back inside the target by the electrostatic field of the sheath, and that it has a regular periodicity $\lambda_{i}\simeq {\lambda}/{4}$, as shown in \fref{fig:Growing-oscillations-of}~a).
These observations will guide us to sketch a simple model for the generation of the ion density perturbation, in which the temporal structure of the fast electron bunches plays an essential role (\sref{Modello}).

\subsection{Comparison with PIC simulations}
 
In this paragraph we show the comparison between the results obtained with the PIC approach and the ones obtained with the Vlasov code. With respect to the Vlasov approach, the well known drawbacks of PIC simulation are the high statistical noise level and the limited resolution of low density regions of the phase space. In order to test the convergence of the PIC simulations, runs were performed for three different numbers of particles per cell per species $N_p=10^{2},10^{3},10^{4}$ and also for two different values of the spatio-temporal resolution $\Delta x=c\Delta t={\lambda}/{10^2}$ and $\Delta x={\lambda}/{10^3}$.

\begin{figure}
\centering{}
\includegraphics[width=0.4\columnwidth]{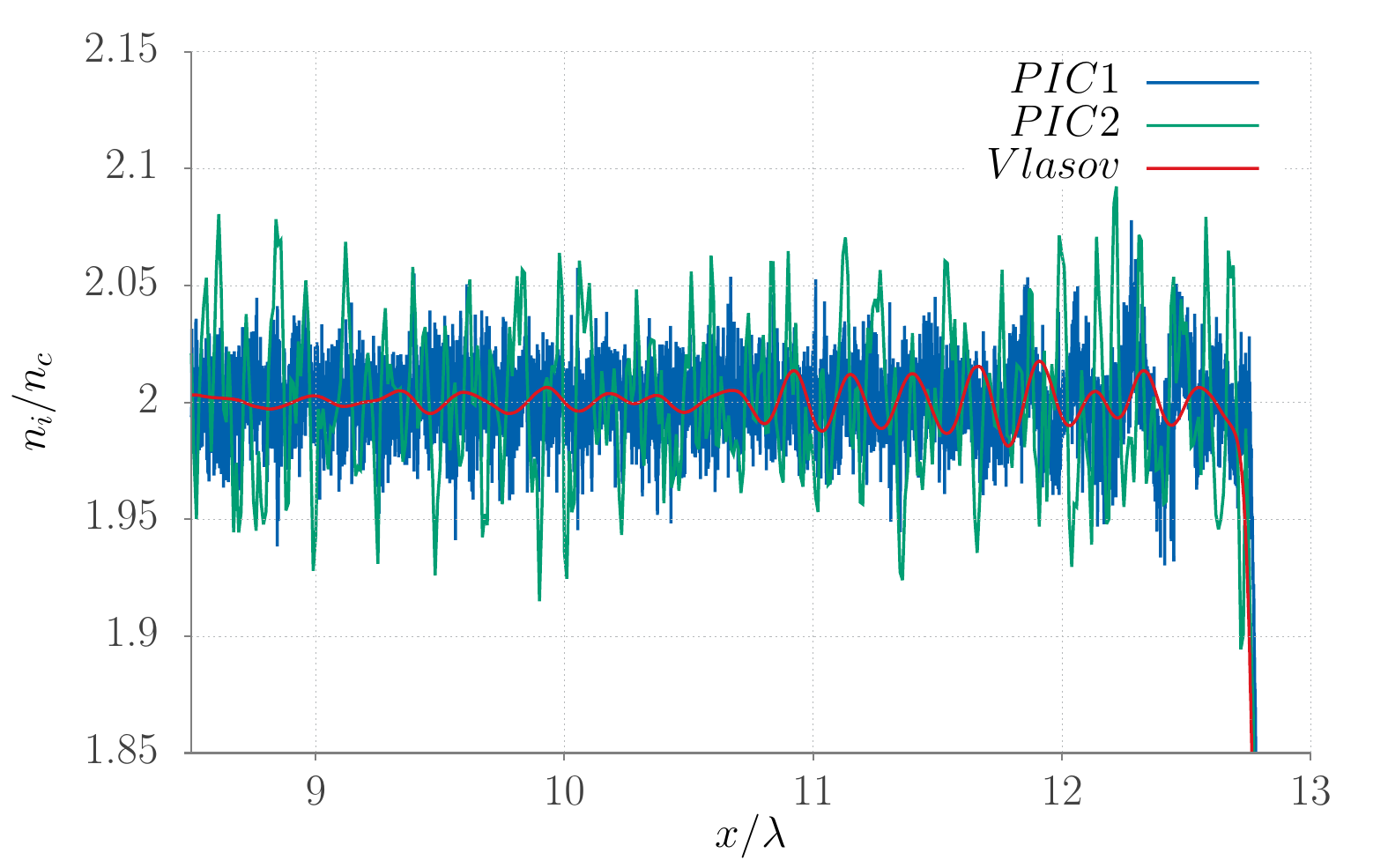}
\includegraphics[width=0.4\columnwidth]{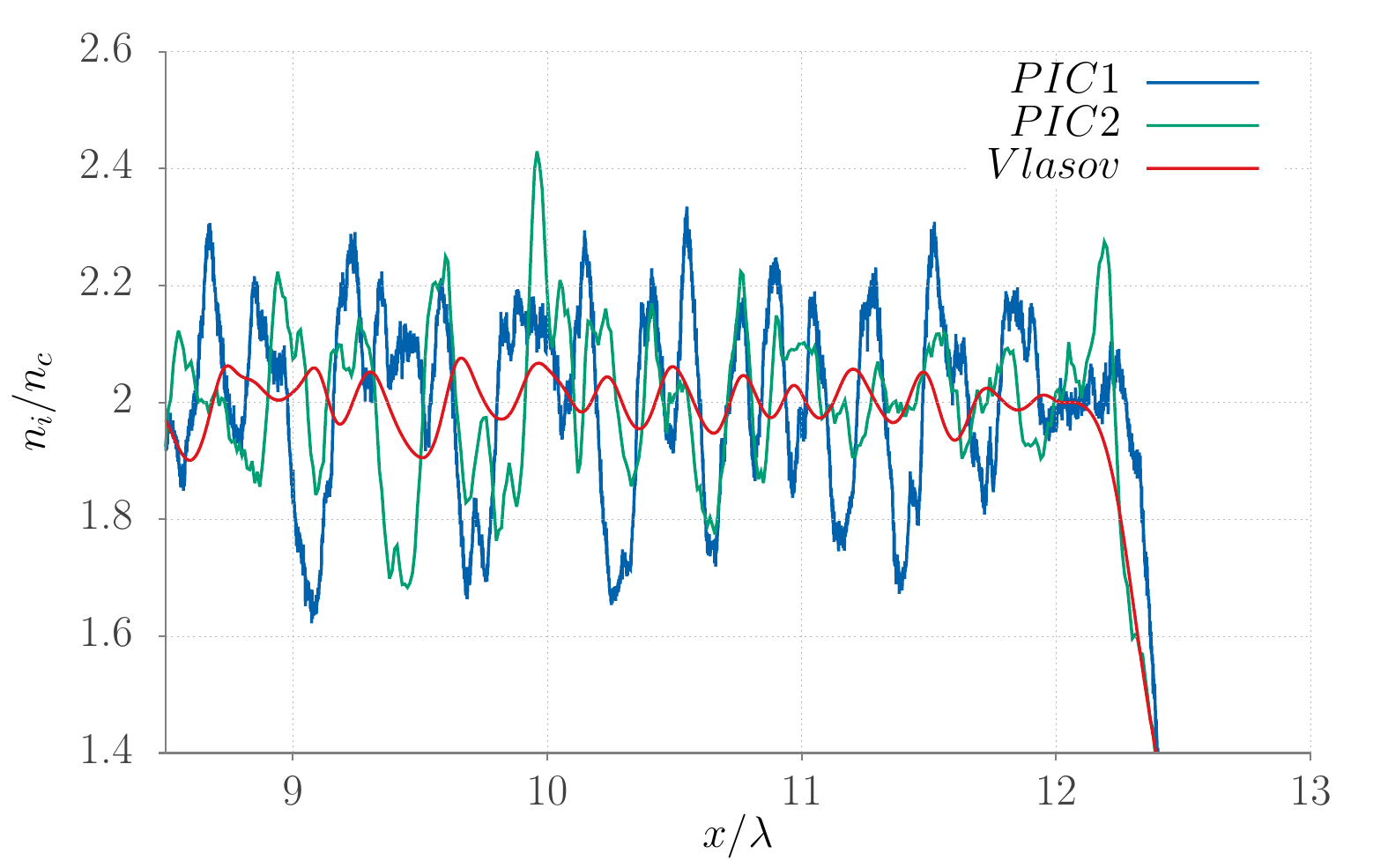}
\caption{\label{fig:Confronto-pic-vlasov}Ion density oscillations at $t=27\: T$
(left frame) and $t=77\: T$ (right frame). PIC1 corresponds to the simulation with $\Delta x={\lambda}/{10^3}$, PIC2 with $\Delta x={\lambda}/{10^2}$.}
\end{figure}

In \fref{fig:Confronto-pic-vlasov} we compare the Vlasov results on the development of ion turbulence with the PIC ones obtained with $N_p=10^{3}$ and the two different choices of $\Delta x$.  Despite the relatively high resolution, the noise level in PIC simulations makes it difficult to clearly characterize the ion perturbation growth and to accurately measure the periodicity of the oscillations, whereas at later times the noise leads to a turbulence in the upstream region with greater amplitude than in the Vlasov case. The difference can be attributed to the higher level of initial noise which acts as a seed for the nonlinear three-wave process generating the turbulence (\sref{Modello}). It may also be noticed that in the Vlasov code the density perturbations are smoothed to some extent by numerical diffusion. 

\begin{figure}
\centering{}
\includegraphics[width=0.4\columnwidth]{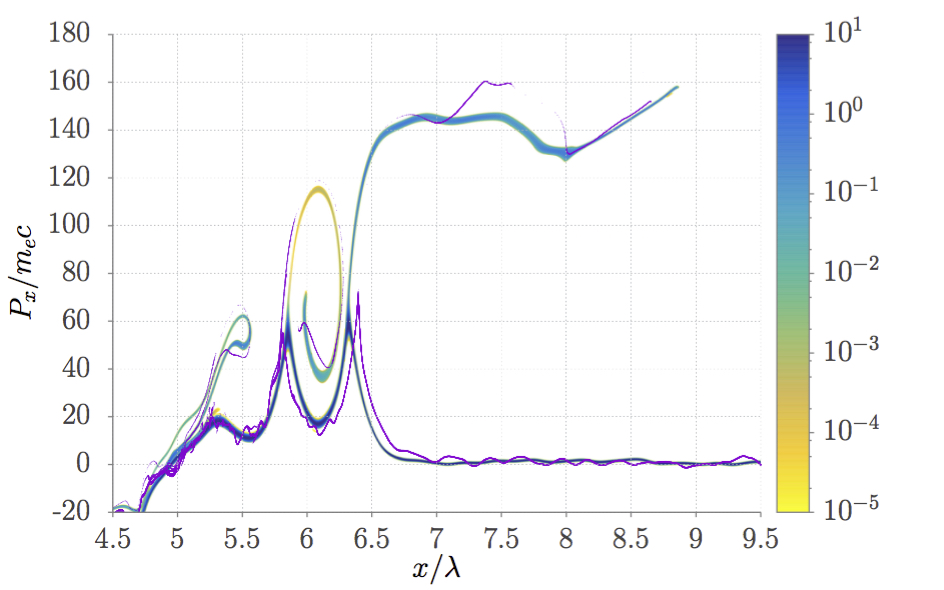}
\caption{\label{fig:PhaseSpace_VP} Ion distribution function at $t=82\: T$ for the Vlasov and PIC (purple dots) simulation with $\Delta x={\lambda}/{10^2}$ (left frame). 
}
\end{figure}

The difference in the amplitude of the ion turbulence affects the ion reflection from the shock front since the broadening of the distribution function in the upstream region is strictly dependent on the characteristics of the turbulence. The different fraction of particles, in the two numerical approaches, lying in a region of the phase space where condition (\ref{eq:minimun_vi}) is fulfilled, may vary the transfer of kinetic energy from the shock front to the accelerated particles. As already discussed in Ref.\cite{macchiPRE12}, a high number of accelerated particles may slow down the shock and at later times may broaden the energy spectrum peak toward lower energies. In \fref{fig:PhaseSpace_VP} the ion phase space from the PIC simulation at $t=82T$ is superimposed to the ion phase space obtained with the Vlasov code. The reflection in the PIC simulation is not as steady as in the Vlasov case and the shock velocity is slightly higher.  Differences are also apparent in the energy spectrum of the accelerated ions (\fref{fig:EnergySpectrum}), with a third spectral peak appearing at $\sim 7$~MeV. Thus, the comparison show that the PIC simulation predicts higher energy and efficiency. 
Finally, a comparison is also shown for the electron phase space in \fref{PSE}. For this latter, the very good agreement is a test of the accuracy of the relativistic splitting scheme for the Vlasov code, including a correction for mass conservation (see \ref{sec:Appendix}).

\section{A model for ion turbulence}
\label{Modello}

In this section we present a simple model which accounts for the generation of ion density oscillations with a spatial periodicity of $\simeq \lambda/4$, where $\lambda$ is the wavelength of the laser pulse. 

In the case of a linearly polarized intense laser pulse incident on an overdense plasma, electrons are accelerated by the magnetic component of the Lorentz force (${\bf J}\times{\bf B}$) and are pushed into the plasma as bunches at a $2\omega$ rate, being $\omega$ the laser frequency. Here we focus on the most energetic electrons propagating at velocities close to $c$ for high enough intensities. These fully relativistic electron bunches maintain their coherence during the propagation. An experimental proof of such coherence comes from the spectrum of the optical transition radiation emitted when the electrons reach the rear boundary of the target\cite{Popescu}. The lower energy part of the distribution function corresponding to velocities $<c$ will lose temporal coherence. In our model we thus assume the coherent, fully relativistic electrons as a population separated from the background electrons which have been heated up forming a Boltzmann distribution with a typical temperature $T_e$, assumed to have a non-relativistic value.

\begin{figure}
\begin{centering}
\includegraphics[width=0.4\textwidth]{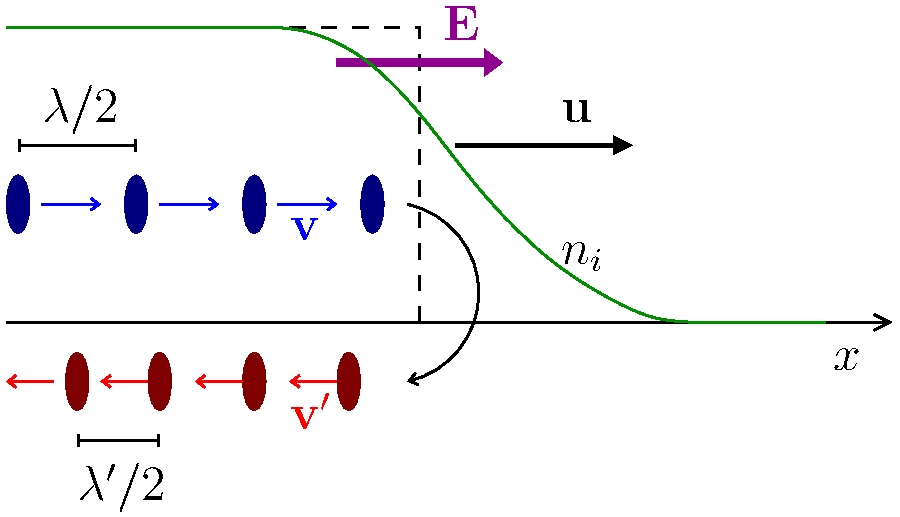}
\caption{\label{fig:Schemino-macchi}
Schematic representation of the relativistic electrons bunches propagating in the positive direction (blue) and of the bunches reflected by the electric field in the expanding sheath (red).}
\par\end{centering}
\end{figure}

Most of the relativistic electrons reflux back inside the plasma because of the electrostatic field that acts as a reflecting wall. Thus, after a time interval which depends on the target size there are two counterstreaming populations of relativistic electrons, both bunched with spatial periodicity $\lambda_{p}\simeq 2\pi c/(2\omega)={\lambda}/{2}$ as it can be noticed in \fref{PSE}. The situation is sketched in \fref{fig:Schemino-macchi}.

In our model we consider the nonlinear beat of the counterpropagating relativistic bunches, which we consider as two pump waves for the system. 
We write the electric field perturbation associated to the bunches as
\begin{eqnarray}
E^{P}(x,t)=\frac{E_{+}}{2}{\mbox{e}}^{ik_{+}x-i\omega_{+}t}+\frac{E_{-}}{2}{\mbox{e}}^{-ik_{-}x-i\omega_{-}t}+\mbox{c.c.} \; ,
\label{eq:pump}
\end{eqnarray}
with $\omega_{\pm} \approx 2\omega$ and $k_{\pm} \approx 2\pi/(\lambda_p)=4\pi/\lambda$. The resulting nonlinear force on the electrons at the beat frequency may be calculated as follows. Starting with the equation of motion in the field (\ref{eq:pump}), 
\begin{eqnarray}
m_e\frac{d^2{x}_e(t)}{dt^2}=-eE^{p}(x_e(t),t) \; , 
\end{eqnarray}
we solve for ${x}_e(t)$ by a standard iterative method. Writing $x_e(t)=x_{0e}(t)+x_{1e}(t)$, we have
\begin{eqnarray}
m_e\frac{d^2{x}_{0e}}{dt^2} \simeq -eE^{p}(x_e(0),t) \; , \quad m_e\frac{d^2{x}_{1e}}{dt^2} \simeq -e{x}_{0e}\partial_x E^{p}(x_e(0),t) \; .
\end{eqnarray}
Thus, in the equation for $x_{1e}(t)$ there is a term at the beating frequency $\Omega=\omega_{+}-\omega_{-}$ and wavevector $K=k_{+}+k_{-}$. The response at such frequency can be described by the nonlinear force
\begin{eqnarray}
f_{\mbox{\tiny NL}}=-\frac{ie^2}{4m_e}\left(\frac{k_{-}}{\omega_{+}^2}E_{+}E_{-}^{*}-\frac{k_{+}}{\omega_{-}^2}E_{-}E_{+}^{*}\right){\mbox{e}}^{iKx-i\Omega t} +\mbox{c.c.}\; .
\label{eq:fNL}
\end{eqnarray}
Notice that $f_{\mbox{\tiny NL}}=0$ for a symmetric situation with $E_{+}=E_{-}$, $\omega_{+}=\omega_{-}$ and $k_{+}=k_{-}$. In our case this would correspond to elastic reflection of the relativistic bunches at the rear plasma boundary. The symmetry is broken because of the loss of energy to the background electrons from the bunches while propagating in the plasma, and at the plasma boundary via the self-generated sheath field as the bouncing electrons enter the vacuum side. In particular, since the background electrons have a temperature $T_e$, the plasma boundary will be in motion with a typical velocity of the order of $c_s=(ZT_e/m_i)^{1/2}$, the ion-acoustic velocity. The reflection of the relativistic electron bunches from the wall moving at velocity $u\simeq c_{s}$ leads to a shift in the reflected frequency that gives different amplitudes between the forward and backward perturbations and a non-zero value of $\Omega=\omega_{+}-\omega_{-} \simeq 2\omega({u}/{c})$ (since $u\ll c$). In particular, $\Omega \sim K c_s$, so that the nonlinear perturbation may couple efficiently to ion-acoustic waves. Such perturbation has wavevector $K=k_{+}+k_{-} \simeq 2k_{\pm}=2\pi/(\lambda/4)$ which agrees with the periodicity observed in the simulations.

The coupling of the nonlinear perturbation at the beating frequency $\Omega$ to ion-acoustic waves may be described by the system of linearized fluid equations
\begin{eqnarray}
\partial_{t}n_{e}+n_0\partial_{x}v_{e}=0 \; , \\
\partial_{t}v_{e}=-\frac{e}{m_e}E-\frac{T_e}{m_e}\frac{\partial_xn_{e}}{n_0}+f_{\mbox{\tiny NL}}\; ,\\
\partial_{x}E=4\pi e (n_{e}-Zn_{i}) \; , \\
\partial_{t}n_{i}+\frac{n_0}{Z}\partial_{x}v_{i}=0 \; ,\\
\partial_{t}v_{i}=\frac{Ze}{m_i}E \; .
\label{eq:full}
\end{eqnarray}
For further simplification we assume that on this low frequency scale the electrons can be considered in a mechanical quasi-equilibrium condition, $v_e\simeq 0$, and quasi-neutrality may be assumed, $n_e \simeq Zn_i$. In this way we obtain for the ion equation of motion
\begin{eqnarray}
\partial_{t}v_{i} \simeq 
-Zc_s^2\frac{\partial_xn_{i}}{n_0}+\frac{Z}{m_i}f_{\mbox{\tiny NL}}\; ,
\end{eqnarray}
so that, using the equation of continuity to eliminate $n_i$ we eventually obtain
\begin{eqnarray}
(\partial_{t}^2-c_s^2\partial_{x}^2)v_{i}=\frac{Z}{m_i}\partial_tf_{\mbox{\tiny NL}} \; ,
\end{eqnarray}
with the solution
\begin{eqnarray}
{v}_{i}=\frac{Z}{m_i}\frac{i\Omega}{(\Omega^{2}-c_{s}^{2}K^{2})}\tilde{f}_{\mbox{\tiny NL}}{\mbox{e}}^{iKx-i\Omega t}+\mbox{c.c.}\; ,
\label{eq:osc}
\end{eqnarray}
where $\tilde{f}_{\mbox{\tiny NL}}$ is the complex amplitude of $f_{\mbox{\tiny NL}}$ [Eq.(\ref{eq:fNL})]. Eq.(\ref{eq:osc}) shows a resonant behavior when the fast electron-driven perturbation frequency is close to that of the ion acoustic waves.

It may be argued that the above described effect is strongly enhanced by the 1D geometry as the reflected electrons are in the same directions as the incoming ones. In a more realistic multi-dimensional geometry the electrons would reflux with some angular spread and the modeling would be more complicated. Nevertheless, our picture shows that in principle the bunched nature of fast electrons, commonly neglected in the modeling of their transport through the plasma, may play an important role in the development of nonlinear effects and instabilities.

\section*{Conclusions }

Generation of laser-driven collisionless shocks and related ion acceleration has been studied by means of numerical Eulerian simulations, based on a corrected splitting scheme for the relativistic Vlasov equation. The high resolution of low-density phase space regions and the absence of numerical noise typical of Vlasov simulations allowed a clear characterization of the shock formation process and the observation of a new mechanism of fast electrons-driven ion turbulence. Hence, the Vlasov simulation approach appears to be a useful tool for the study of collisionless shock acceleration physics and it is being used already by other groups \cite{wettervikXXX15}.
  
In the regime we investigated, the shocks are driven by the radiation pressure action and related wavebreaking in the ion density profile. Ion reflection from the shock front occur also for initially cold ions because of the velocity spread generated by the development of ion turbulence in the upstream region. Particle-in-cell simulations show a higher amplitude for the turbulent oscillations which leads to differences in the spectrum of accelerated ions with respect to Vlasov simulations. A simple model in which the ion turbulence is driven by beats induced by pulsed, coherent, relativistic electron bunches has been introduced.

\appendix

\section{Mass conservation in the splitting scheme}
 \label{sec:Appendix}

In our code the so called Time Splitting Scheme \cite{Cheng1976330} is exploited in order to perform the numerical integration of the Vlasov equation.
The scheme has been widely used and benchmarked in the electrostatic non relativistic case; in order to calculate the evolution of the distribution function
it treats separately in the Vlasov equation the convective terms in the $x$ direction and the one along the momentum axis,
obtaining a scheme accurate up to the second order.

In the fully relativistic electromagnetic case, it has been demonstrated that the scheme does not conserve the particle density \cite{Huot2003512}. Indeed in our case the Vlasov equation can be splitted as
\begin{eqnarray}
\partial_t f_{1}+\partial_{x}\left(\frac{p_{x}}{m\gamma}f_{1}\right)=-f_{1}\partial_{x}\left(\frac{p_{x}}{m\gamma}\right)\\
\partial_t f_{2}+\partial_{p_{x}}\left(F_{L,x}f_{2}\right)=-f_{2}\partial_{p_{x}}(-\frac{q^{2}}{2m\gamma}\frac{\partial\mid\vec{A}{}_{\perp}\mid^{2}}{\partial x})
\label{eq:Splitted}
\end{eqnarray}
where $F_{L,x}$ is Lorentz force along the $x-$axis and $\gamma=({1+[{p^2_x+q^2A^2_{\perp}(x)}]/{m^2}})^{1/2}$.

The presence of the relativistic $\gamma$ factor leads to the appearance of extra terms on the r.h.s., which are vanishing in the electrostatic non-relativistic case for which $\gamma=1$ \cite{Huot2003512}. Applying the method presented in Ref.\cite{Filbet2001166} to calculate the evolution of the distribution function, a cumulative systematic error would be introduced at each time step, resulting in a poor conservation of the charge density for each species. Without applying any correction, an unphysical loss of charge is found in the region where the electromagnetic effects are important (i.e. in correspondence of the laser pulse, where $\vec{A}_{\perp}\neq0$). On the other hand, in the region where the transverse vector potential vanishes the system is automatically reduced to the electrostatic case.

\begin{figure}
\centering{}
\includegraphics[width=0.5\columnwidth]{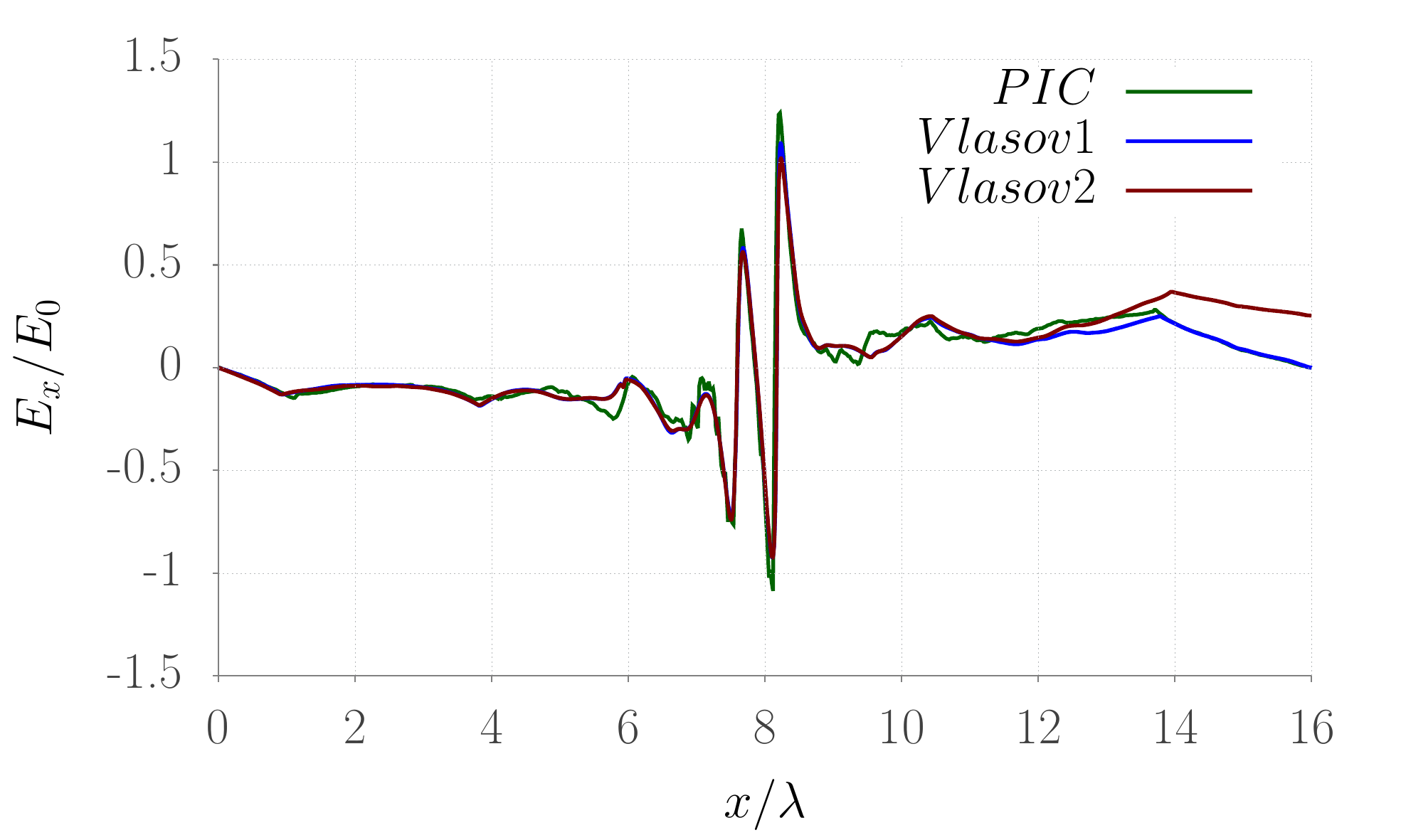}
\caption{\label{fig:Toppa}
Comparison of the longitudinal electric field at
$t=75\: T$ for a Vlasov simulation carried out with (blue line) and
without (red line) the correction and a PIC (green line) simulation. }
\end{figure}

In order to ensure the total particle mass conservation the quantities on the r.h.s. of Eq.(\ref{eq:Splitted}) are considered as local source terms. Therefore these terms are integrated at each time step and the corresponding density is reintroduced in the distribution function reproducing in the momentum space the distribution of the remaining particles. As an example, in Fig.\ref{fig:Toppa} we report the comparison between a simulation carried out without the correction and a corrected one, running with the same parameters. We also show a PIC simulation that provides a test of the correctness in the relativistic and electromagnetic regime of the correction developed in our Vlasov code. 

Fig.\ref{fig:Toppa} shows the longitudinal electric field associated with the ion density peak propagating in the overdense plasma. If the correction is not applied, the loss in the electron density causes the plasma to be non exactly neutral and an electric field arises at the right boundary. Reintroducing the lost electron mass, the $E_{x}$ field vanishes at the end of the box.

\ack
This work has been supported by MIUR, Italy, via the PRIN project "Laser-driven shock waves" (grant 2012AY5LEL\_002)

\section*{References}


\end{document}